# Spin-orbit field in a physically-defined p-type MOS silicon double quantum dot

Marian Marx[1,2*], Jun Yoneda[2,3], Ángel Gutiérrez-Rubio[2], Peter Stano[1,2,4], Tomohiro Otsuka[2,5,6,7], Kenta Takeda[2], Sen Li[2], Yu Yamaoka[8], Takashi Nakajima[2], Akito Noiri[2], Daniel Loss[2,9], Tetsuo Kodera[8] and Seigo Tarucha[1,2*]

[1]*Dept. of Applied Physics, The University of Tokyo, Bunkyo-ku, Tokyo 113-8656, Japan*

[2]*RIKEN Center for Emergent Matter Science (CEMS), Wako, Saitama 351-0198*

[3]*School of Electrical Engineering and Telecommunications, University of New South Wales, Sydney, New South Wales 2052, Australia*

[4] *Institute of Physics, Slovak Academy of Science, 845 11 Bratislava, Slovakia*

[5]*Research Institute of Electrical Communication, Tohoku University, 2-1-1 Katahira, Aoba-ku, Sendai 980-8577*

[6]*Japan JST, PRESTO, 4-1-8 Honcho, Kawaguchi, Saitama 332-0012, Japan*

[7]*Center for Spintronics Research Network, Tohoku University, 2-1-1 Katahira, Aoba-ku, Sendai 980-8577, Japan*

[8]*Department of Electrical and Electronic Engineering, Tokyo Institute of Technology, Meguro, Tokyo 152-8552, Japan*

[9]*Department of Physics, University of Basel, Klingelbergstrasse 82, CH-4056 Basel, Switzerland*

*e-mail: marian.marx@riken.jp; tarucha@riken.jp

We experimentally and theoretically investigate the spin-orbit (SO) field in a physically-defined, p-type metal-oxide-semiconductor double quantum dot in silicon. We measure the magnetic-field dependence of the leakage current through the double dot in the Pauli spin blockade. A finite magnetic field lifts the blockade, with the lifting least effective when the external and SO fields are parallel. In this way, we find that the spin-flip of a tunneling hole is due to a SO field pointing perpendicular to the double dot axis and almost fully out of the quantum-well plane. We augment the measurements by a derivation of SO terms using group-symmetric representations theory. It predicts that without in-plane electric fields (a quantum well case), the SO field would be mostly within the plane, dominated by a sum of a Rashba- and a Dresselhaus-like term. We, therefore, interpret the observed SO field as originated in the electric fields with substantial in-plane components.

## I. INTRODUCTION: SPIN-ORBIT INTERACTION IN P-TYPE MOS SILICON DQD

Among targets of the research with semiconductor quantum dots is the implementation of spin qubits [1–6]. Silicon devices are particularly appealing since only 5% of the nuclei carry spins [7–11], which can be further reduced by isotopic purification to enhance electron spin coherence [12–14]. Hole spins in Si should be naturally further isolated from the nuclear spin noise due to their p-orbital nature, making them a promising qubit system with potentially long coherence times [15–18]. Moreover, the spin-orbit (SO) interaction, an efficient coupling mechanism between spin and electric fields [19], is stronger compared to electrons [20–22]. It can be exploited for spin manipulation via electric dipole spin resonance (EDSR) [23–29] so that an external source of spin-electric coupling (such as micromagnets [30–32]) is not needed. This simplifies the device design and makes it more compatible with the standard complementary metal-oxide-semiconductor (CMOS) fabrication and thus benefits the upscaling and compatibility with classical electronics.

On the other hand, a large SO interaction beneficial for qubit controllability might also become a major decoherence source [33–35]. Similarly, it has adverse effect on the spin readout via the Pauli spin blockade (PSB) [36–43]: the PSB gets lifted by relatively small magnetic fields of the order of tens of mT, whereas a field exceeding 100 mT is desirable to raise the Zeeman splitting reliably above typical thermal energies. Concerning spin qubits in MOS devices [9,12,28,37,38,42,44–46], anisotropic g-factors were studied for electrons in Ref. [47] and for holes by Liles, *et al.* [48] by pulsed DC measurement and by A.Crippa, *et al.* [49] using EDSR. However, while the g-factor is related to the SO interaction, the relationship is far from straightforward [50]. In this work we therefore target directly the SO field, by investigating the lifting of the PSB in a physically defined pMOS Si dot. Our device combines the advantages of the industry-standard CMOS fabrication with additional tunability provided by a separate top gate and plunger gates. Although other mechanisms such as co-tunneling, spin relaxation, or effects from nuclear spins can also contribute, the primary mechanism of PSB lifting of holes in Si is the SO interaction: the combination of the SO and external magnetic fields allows the hole to effectively flip its spin upon interdot tunneling, which lifts the blockade.

The essential difference of our approach to the majority of recent works is that we do not assume, a priori, a specific form for the SO interactions. They are usually taken as a combination of Rashba and Dresselhaus terms [28], or just one of them [51,52]. Instead, an important part of our investigations is a theoretical analysis which relies on symmetry, considering the crystal, the interface, and the quantum dot together. This approach goes along pioneering works that showed that abrupt potential changes at interfaces can result in terms contradicting the conventional knowledge, such as a "Dresselhaus" term in material with bulk inversion symmetry or a "Rashba" term in a macroscopically symmetric quantum well [53–55]. Correspondingly, we find terms that are generated by electric fields, but cannot be written simply as $\vec{E} \cdot (\vec{k} \times \vec{\sigma})$, a generic "Rashba" term (see Section IV).

## II. DEVICE DESCRIPTION: A PHYSICALLY-DEFINED DOUBLE QUANTUM DOT

The double quantum dot (DQD) used in the experiment is made on silicon-on-insulator [Fig. 1(a)] with the wafer-surface normal along $\hat{y} = [110]$. It is the sample used in Ref. [46]. A buried oxide separates the Si substrate from the Si quantum well layer [orange in Fig. 1(a)], where the DQD is etched. The DQD is aligned with $\hat{x} = [1\bar{1}0]$ and the in-plane direction perpendicular to it is $\hat{z} = [001]$. On top of this a 50nm thick gate oxide is grown which is topped by highly doped Si (poly Si) serving as a global accumulation gate (top gate). It covers the whole depicted part of the device and induces the holes. Further details of the device fabrication can be found in Ref. [56]. The dot occupancy is tuned by plunger gates (SGL and SGR). The device also contains a charge sensor (not depicted) [56], here used only to estimate the hole occupancy. The occupancy is tuned to the order of 10 holes, as counted from the charge stability diagram, to have a sufficient transport signal.

We examine the PSB leakage current through the DQD at zero level detuning ($\varepsilon = 0$) while changing the magnitude of the external magnetic field $\vec{B}_{\text{ext}}$ (shown as a green arrow) applied in various directions [Footnote1]. Fig. 1(b) shows the angle coordinates that we use: $\theta$ is the angle between $z$-axis and $\vec{B}_{\text{ext}}$ and $\phi$ is the angle between $x$-axis and the projection of $\vec{B}_{\text{ext}}$ in the $x$-$y$ plane. As explained below in section IV,

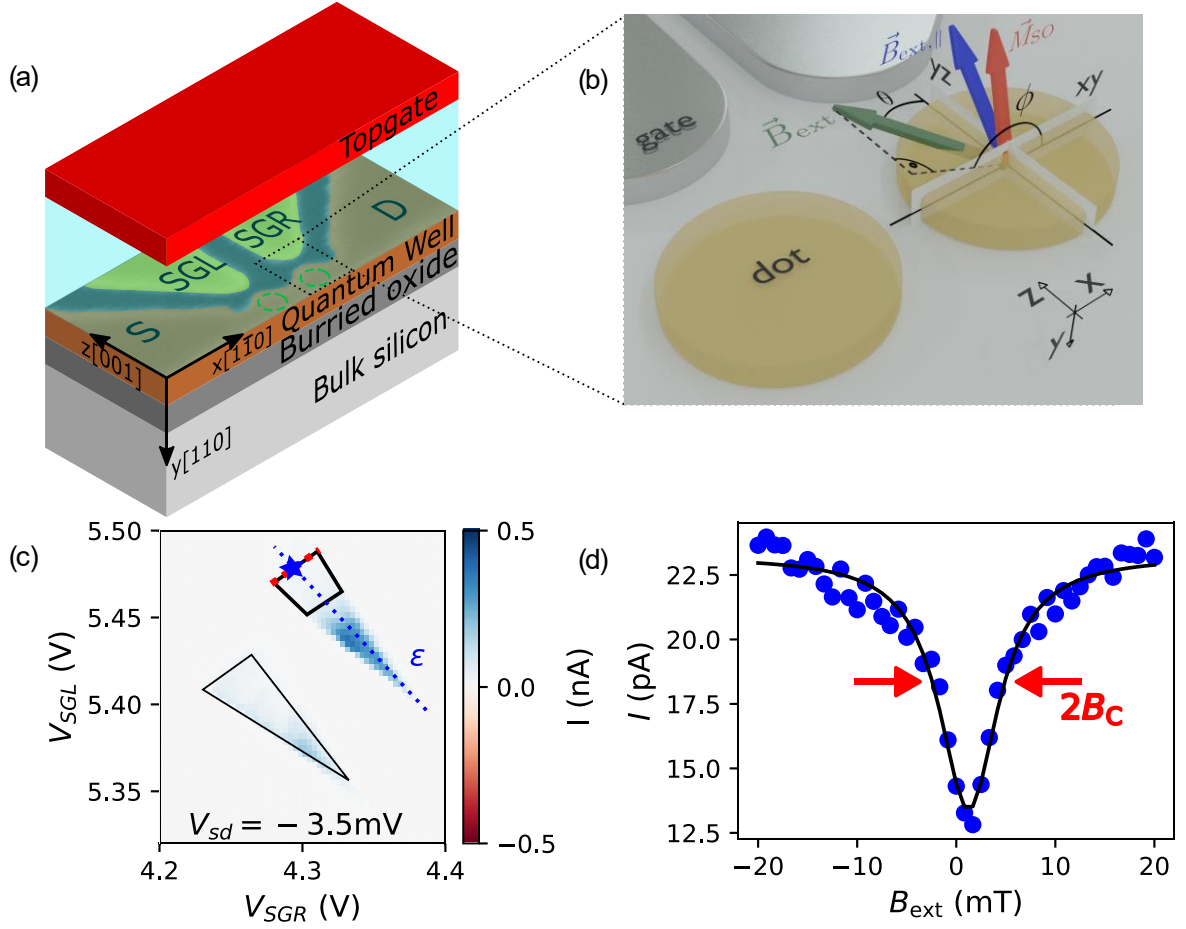

**Figure 1.(a)** Schematic of the device. **(b)** Definition of the angles and presentation of the resulting vectors for the SO field $\vec{M}_{SO}$ and the external field for the largest resilience to lifting the PSB $\vec{B}_{ext,\parallel}$. **(c)** Current through a DQD as a function of the voltages applied on SGR and SGL with a top-gate voltage of $V_{TG} = -4.6V$. The red dotted line marks the baseline $\epsilon = 0$ of the bias triangle and the blue star the measurement configuration for (d). **(d)** Current measured at the position marked with a blue star in (c) as a function of the modulus of $\vec{B}_{ext}$ applied along $\hat{x}$. The peak-center offset from zero by $0.77\,mT$ is likely due to a residual magnetization of the magnet.

the effects of the SO interaction are grasped by the following Hamiltonian [19]:

$$H_{SO}^{PSB} \equiv \mu_B \vec{M}_{SO}(\vec{k}_{12}) \cdot \vec{\sigma}. \tag{1}$$

With this formula, we assign the DQD an effective SO field $\vec{M}_{SO}$, where $\mu_B$ is the Bohr's magneton, $\vec{k}_{12}$ the momentum when tunneling from dot 1 to dot 2, and $\vec{\sigma}$ the Pauli vector representing the effective (pseudo) spin of the hole ground state. As will be explained below, the SO field in Eq. (1) is generated by electric fields along the growth direction (related to the quantum-well interfaces and the top gate) as well as in-plane (related to the in-plane confinement due to etching and the side gate). The SO field influences the PSB by inducing spin-flipping interdot tunneling. Our approach relies on the fact that the SO effects are generically anisotropic, depending on the device geometry and the orientation with respect to the crystal axes. The anisotropy has been instrumental in detecting, identifying, and tuning the SO effects [8,57–65]. Accordingly, we determine the direction of $\vec{M}_{SO}$ from the dependence of the PSB leakage current on the magnetic-field direction [Footnote2].

### III. DETERMINATION OF THE SPATIAL DIRECTION OF THE SPIN-ORBIT FIELD

In Fig. 1(c) we plot the bias triangle at the transition between *(m+1,n+1)* and *(m,n+2)* states with a source-drain bias $V_{sd} = -3.5 \text{mV}$. (See Ref. [56] for a charge stability diagram of a nominally identical device measured for a much larger range of side gate voltages). As in the lower bias triangle of Fig.1(c) the blocked region is less clear, we focus on the upper triangle. The trapezoidal region [outlined in Fig1(c)] of suppressed current near the base line (the level detuning $\varepsilon = 0$, shown by the red dashed line) indicates the PSB. In Fig. 1(d), we plot the leakage current $I$ as a function of the external magnetic field modulus $B_{\text{ext}}(= \|\vec{B}_{\text{ext}}\|)$ for a fixed direction given by $\left(\theta = \frac{\pi}{2}, \phi = 0\right)$. The gate configuration used in this measurement is denoted by a star in Fig. 1(c) and is fixed for all further measurements. The current suppression at zero field and the lifting of PSB at finite field indicates a considerable SO interaction. The crossover scale is defined as the width of the current dip, denoted by $B_C$. For $B_{\text{ext}} \ll B_C$ the two-spin eigenstates are dominated by the SO energy, whereas with increasing $B_{\text{ext}}$ the competition of the Zeeman energy and the SO energy lifts the PSB until for $B_{\text{ext}} \gg B_C$ the two-spin eigenstates are dominated by the Zeeman energy. Following Ref. [36], we fit the $I - B_{\text{ext}}$ curve with a Lorentzian,

$$I = I_0 + \delta I \frac{B_C^2}{B_{\text{ext}}^2 + B_C^2} . \quad (2)$$

In this formula, $B_C$, $\delta I$, and $I_0$ are fit parameters. We use this formula below to fit all traces such as the one in Fig. 1(d).

We first investigate the in-plane magnetic-field-direction dependence of the crossover field $B_C$. To this end, we measure the $I - B_{\text{ext}}$ traces at different $\theta$ and a fixed $\phi = 0$. The traces, plotted in Fig. 2(a), show a dip as predicted by Eq. (2). They also become flatter and slightly asymmetric near $\theta \approx 0.9\pi$. As here they fit the Lorentzian less well (though that is difficult to see at the figure resolution), there is a larger error in the extracted $B_C$ [visible in the error bars in Fig. 2(b)]. The shape distortion might be due to competing blockade lifting mechanisms, neglected in our theoretical model, which are visible where the SO effects become suppressed. The distortion effects are much stronger for an out-of-plane magnetic field (see below).

Assuming that the spin-conserving tunneling rates are not influenced by the modest magnetic field $\vec{B}_{\text{ext}}$, Ref. [36] predicts that $B_C$ is inversely proportional to the outer product of the spin-orbit field and the external magnetic field. We generalize this relation for a non-isotropic g-tensor $\bar{\bar{g}}$ [Footnote3] into the following form

$$B_C \propto \frac{\|\vec{M}_{SO}\|\|\vec{B}_{\text{ext}}\|}{\|\vec{M}_{SO} \times (\bar{\bar{g}} \cdot \vec{B}_{\text{ext}})\|} . \quad (3)$$

From symmetry analysis we conclude that the g-tensor for holes in our double dot is well approximated by (see App. A)

$$\bar{\bar{g}} \approx \begin{pmatrix} g_{xx} & 0 & 0 \\ 0 & g_{yy} & g_{yz} \\ 0 & g_{yz} & g_{zz} \end{pmatrix}. \quad (4)$$

In Ref. [46] we determined that $|g_{xx}| \approx 2|g_{zz}|$. Even though the symmetry analysis cannot determine the magnitudes of the elements, one expects that the off-diagonal element in Eq. (4) is smaller than the diagonal ones. In addition, as described below, we have found that the fits are weakly sensitive to the value of the off-diagonal element $g_{yz}$. We therefore neglect it and use [Footnote4]

$$\bar{\bar{g}} \propto \begin{pmatrix} 2 & 0 & 0 \\ 0 & \gamma & 0 \\ 0 & 0 & 1 \end{pmatrix}, \tag{5}$$

with $\gamma = \frac{g_{yy}}{g_{zz}}$, as the g-tensor in Eq. (3), insensitive to an overall scale.

With the model specified by Eqs. (2), (3), and (5) we fit the data plotted in Fig. 2(b) to Eq. (3), aiming to extract the direction of the vector $\vec{M}_{SO}$. We parameterize the latter by two angles, $(\phi_{SO}, \theta_{SO})$. These angles are the only fit parameters, since the g-tensor matrix element $\gamma$ does not enter Eq. (2) for an in-plane magnetic field. The best fit is plotted as the orange line in Fig. 2(b) and corresponds to $(\pm\phi_{SO}, \theta_{SO}) = (0.342\pi, 0.391\pi) = (61.6°, 70.4°)$. To characterize the error of this estimate we use the $\chi^2$ statistics (see App. B for the definition) as the figure of merit of the fit. We plot it in Fig. 2(c) showing, in the space of fitting parameters, the confidence region corresponding to three standard deviations (App. B explains the conversion from $\chi^2$ to standard deviations). That figure also shows two discrete symmetries: first, the fit figure of merit (the $\chi^2$ statistics) is invariant to the inversion of the SO field. This property is clear from Eq. (3) and in our parametrization corresponds to $(\phi_{SO} \to \pi + \phi_{SO}, \theta_{SO} \to \pi - \theta_{SO})$. Second, with a diagonal g-tensor and the magnetic field in the plane, the fit is invariant with respect to inverting the $y$ component of the SO field, corresponding to $(\phi_{SO} \to 2\pi - \phi_{SO}, \theta_{SO} \to \theta_{SO})$. The best estimate for the SO direction from the in-plane data is $\vec{M}_{SO} \propto \pm(0.45, \pm 0.83, 0.34)$ as the one fitting best, but the errors of the components are large: for example, a vector fully aligned with the $y$-axis is within the confidence region plotted in Fig. 2(d).

To increase the estimate precision and resolve the sign ambiguity, we repeat the above procedure for an out-of-plane field, varying it in the $y$-$z$ plane. Figs. 2(d)-(e) show the corresponding traces, varying $\theta$ for a fixed $\phi = 0.5\pi$. While the majority of the traces still fit to Eq. (2) well, we also observe strong deviations. Especially around $\theta \approx 0.30\pi$ in Fig. 2(d) and around $\theta \approx 0.85\pi$ in Fig. 2(e) the traces do not fit a single Lorentzian: strongly asymmetric peaks develop around the center. While the disappearance of the (main) dip is predicted in Ref. [36] when $\bar{\bar{g}} \cdot \vec{B}_{ext}$ and $\vec{M}_{SO}$ are aligned, the fact that we observe strong deviations from Lorentzian along two almost perpendicular directions is puzzling. We proceed without having an explanation for the origin of this structure and since a strong deviation from Eq. (2) precludes assigning a meaningful value of $B_C$, we omit the anomalous traces from considerations. The omitted traces are denoted by plotting the Lorentzian fits in Fig. 2(d)-(e) in black dashed curves. The corresponding values of $B_C$ (obtained by formally insisting on a Lorentzian fit) are plotted with open symbols in Fig. 2(f) for completeness.

Adding the out-of-plane data to the in-plane ones, we fit the whole set of extracted values of $B_C$ to Eq. (3). The best fit gives $(\phi_{SO}, \theta_{SO}) = (0.496\pi, 0.432\pi) = (89.3°, 77.8°)$ and $\gamma = 3.9$, and the $B_C$ predicted using these values is plotted as a function of $\theta$ for $\phi = 0$ in Fig. 2(b) and for $\phi = 0.5\pi$ in Fig. 2(f) as a blue line. The $\chi^2$ statistics showing the confidence region within $3\sigma$ around the best fit, projected on the plane of $\phi_{SO}$ and $\theta_{SO}$ [the analog of Fig. 2(c)], is plotted in Fig. 2(g). From the same $\chi^2$ statistics we also conclude that the data do not give a useful estimate for the out-of-plane g-factor: the $3\sigma$ region covers the range $\gamma \in [3, 25]$

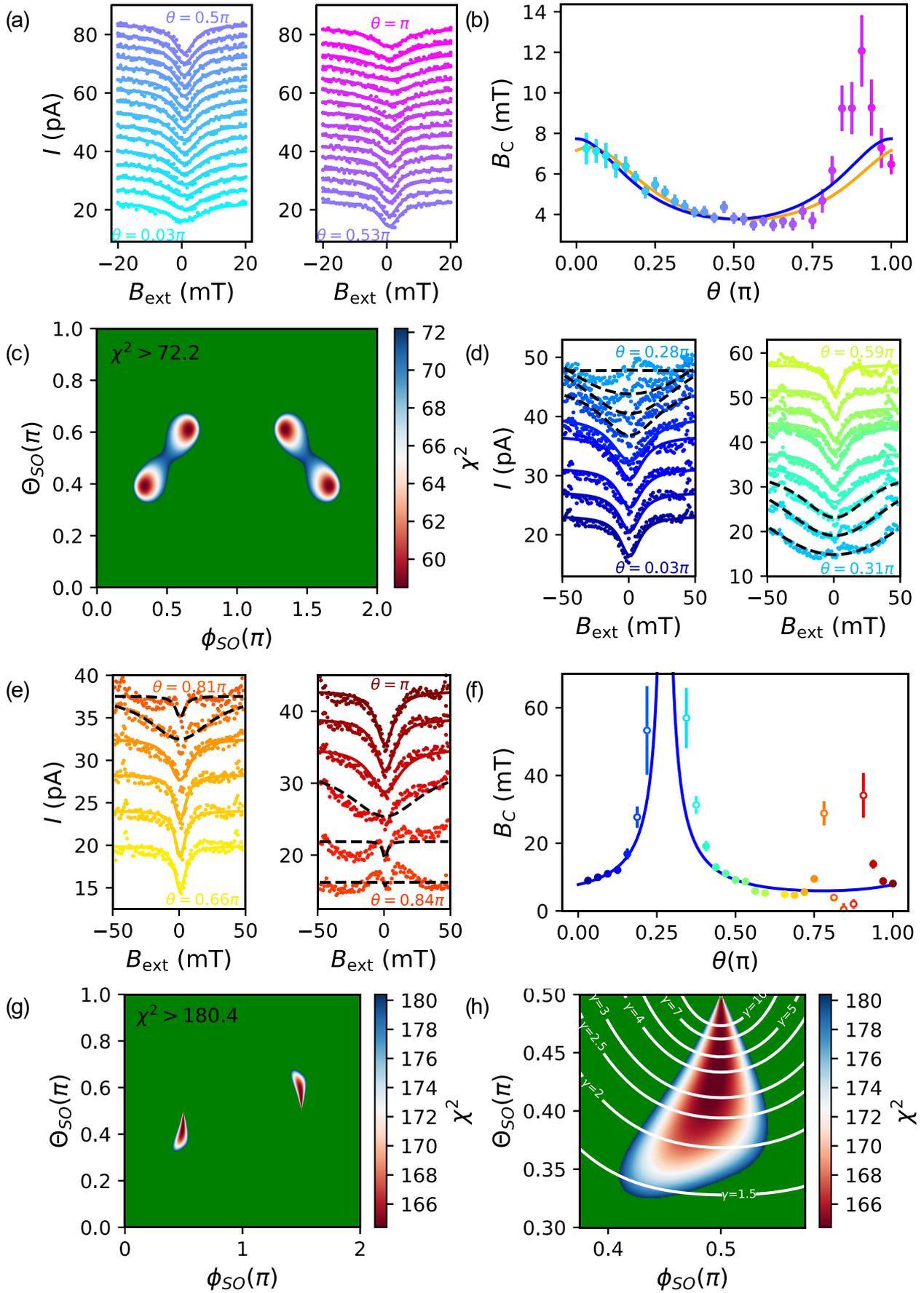

**Figure 2 (a) Magnetic-field sweeps (each offset by 4pA) for $\phi = 0$ and $\theta$ in steps of $\frac{\pi}{32}$. (b) $B_C$ extracted by fitting datasets in (a) to Eq. (2), resulting in a point per dataset plotted in matching colors. The orange line is a fit to Eq. (3) for $(\phi_{SO}, \theta_{SO}) = (0.342\pi, 0.391\pi)$. (c) The fit $\chi^2$ statistics as a function of the fitting parameters, with the range 58-72 corresponding to approximately $3\sigma$ confidence interval. (d,e) Magnetic-field sweeps (offset by 4pA) for $\phi = 0.5\pi$ and $\theta$ in steps of $\frac{\pi}{32}$. (f) $B_C$ fitted using Eq. (2), with one point per dataset in (d,e) plotted in matching colors. [Points in black represent "failed" fits plotted in black dashed curves in (d,e). Out of these, we omitted plotting points with $\theta = 0.25\pi$ and $\theta = 0.31\pi$ which are out of the plotted area, and with $\theta = 0.28\pi$ the error of which is larger than the y-axis range shown.] The curve is a fit to Eq. (3) considering the in- and out-of-plane data together, resulting in the best fit for $(\phi_{SO}, \theta_{SO}) = (0.496\pi, 0.432\pi)$. (g) The $\chi^2$ statistics of the fit to Eq. (3) marginalized over the g-factor, that is $\min_\gamma \chi^2[\phi_{SO}, \theta_{SO}, \gamma]$ with the color bar covering the range corresponding to a $3\sigma$ confidence interval. (h) A zoom of the left minimum shown in (g). Here we additionally show contours of $\gamma$ for which $\chi^2[\phi_{SO}, \theta_{SO}, \gamma]$ is minimal for fixed values of $\phi_{SO}$ and $\theta_{SO}$.**

[see Fig. 3(h)], in addition to the overall scale of the matrix in Eq. (5) being unknown. More importantly, adding the out-of-plane data does resolve the sign ambiguity in the $y$-component of $\vec{B}_{SO}$ and shifts the best fit somewhat closer to $\phi_{SO} = \pm\frac{\pi}{2}$, as can be seen comparing Fig. 2(b) to Fig. 2(g). The error on the angle $\phi_{SO}$, which parametrizes the deflection from this plane, is somewhat smaller than the error of $\theta_{SO}$, the angle within this plane. For illustration, projecting a $1\sigma$ confidence region on the vertical and horizontal axes in Fig. 2(g) gives error estimates $\delta\phi_{SO} \in (-0.045\pi, +0.020\pi) = (-8.1°, 3.6°)$, $\delta\theta_{SO} \in (-0.072\pi, +0.066\pi) = (13.0°, 11.9°)$. We, therefore, conclude that the best estimate for the spin-orbit field direction is $\vec{M}_{SO} \propto \pm(0.01, 0.98, 0.21)$ [the red arrow in Fig. 1(b)], that is within the fitting uncertainty a field almost fully within the $y$-$z$ plane. In addition, our analysis suggests that the biggest component of the g tensor is the out-of-plane one and that the external field direction for the largest resilience to the lifting of the PSB is $\vec{B}_{ext,\|} \propto \pm(0.01, 0.77, 0.64)$ corresponding to $(\phi, \theta) = (0.494\pi, 0.278\pi) = (88.9°, 50.1°)$ [the blue arrow in Fig. 1(b)]. This finishes our experimental investigations of the spin-orbit field and we now turn to its theoretical analysis.

### IV. THE SPIN-ORBIT INTERACTION DERIVED FROM SYMMETRY REPRESENTATIONS

In this section, we derive an effective spin-orbit Hamiltonian theoretically, using symmetry analysis. Hence, we make no other assumptions than the ones imposed by the crystal symmetry of silicon and the device geometry. As for the former, the effective SO Hamiltonian for bulk silicon inherits the symmetries of the octahedral point group $O_h = T_d \times \{I\}$, where $T_d$ and $\{I\}$ are the tetrahedral and inversion groups, respectively. This decomposition allows to reduce the point group to the simpler $T_d$ (the case of GaAs) with the condition that only terms that preserve inversion invariance are allowed [66].

Departing from there, we apply the constraints imposed by the geometry of our experimental setup, which encompass: (i) Interested in a low-energy description, we keep only terms linear in the hole wavevector components $k_i$, $i \in \{x, y, z\}$. (ii) The holes are bound to within the quantum well ($x$-$z$ plane), thus $k_y = 0$. (iii) The quantum well confinement splits the four-fold degeneracy of the valence band: At zero magnetic field, the ground state is a Kramers doublet, described by effective spin ½ operators $\{\sigma_x, \sigma_y, \sigma_z\}$ (Pauli matrices). We derive a Hamiltonian restricted to this two-dimensional subspace. (iv) There are strong electric fields along the $y$ and $z$-direction. No appreciable field is expected along the dot axis, $E_x \approx 0$. (v) The allowed form of the interactions depends on the number of monoatomic layers in the quantum well and the symmetry of its interfaces. Here we give results for the highest symmetry possible ($D_{2h}$) which corresponds to a quantum well with the inversion symmetry (disregarding the electric fields) and the two interfaces "averaged" by monoatomic fluctuations [67] on the scale of the quantum dot (see App. A for details). Depending on the interface and quantum well properties, we identify four additional possibilities (symmetry groups), the results for which are given in App. A.

With these assumptions, in App. A we obtain the following leading-order spin-orbit Hamiltonian for holes in a Si/SiO$_2$ quantum well grown along $\hat{y} = [110]$:

$$H_{SO}^{D_{2h}} = c_1 E_y k_x \sigma_z + c_2 E_y k_z \sigma_x + c_3 E_z k_x \sigma_y + c_4 E_x k_z \sigma_y. \tag{6}$$

Here, $c_1$, $c_2$, $c_3$ and $c_4$ are prefactors that the symmetry analysis cannot specify. Before we apply this Hamiltonian to the double dot experiment, we illustrate its content considering a two-dimensional hole gas of the quantum well. To this end, we follow the standard procedure and keep only the in-plane momenta $k_x, k_z$, and the electric field across the quantum well $E_y$. Putting the remaining momentum and field components to zero we get

$$H_{SO}^{[110]} = E_y (c_1 k_x \sigma_z + c_2 k_z \sigma_x). \tag{7}$$

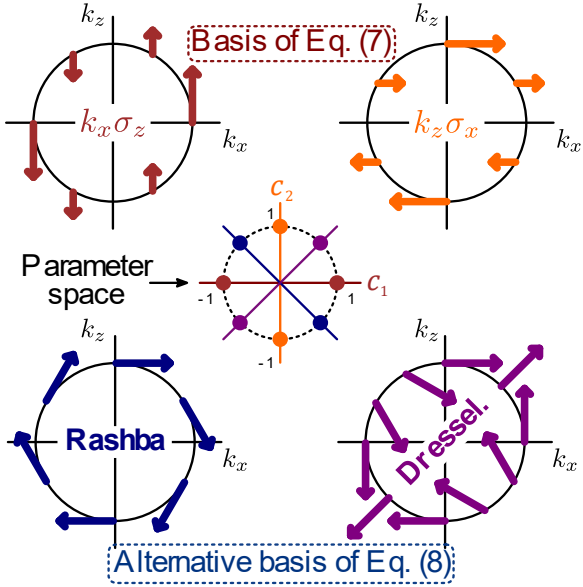

Figure 3. Sketch of the SO terms. The upper diagrams show the basis of the SO Hamiltonian given by the first (left) and second (right) terms of Eq. (7). The lower two show the alternative basis of Rashba-like and Dresselhaus-like combinations in Eq. (8). The parameter space (in arbitrary units) is outlined in the center, sharing the color code with the outlines and showing the relation between both bases, namely a $\frac{\pi}{4}$ rotation.

To connect to the spin-orbit interactions forms well known from 2DEGs, we cast this Hamiltonian in terms of the Rashba-like and Dresselhaus-like combinations [Footnote5],

$$H_{SO}^{[110]} = \frac{c_2 - c_1}{2} E_y (k_z \sigma_x - k_x \sigma_z) + \frac{c_1 + c_2}{2} E_y (k_x \sigma_z + k_z \sigma_x). \tag{8}$$

Figure 3 depicts the relation of the two linearly independent terms in both bases.

After the detour to the case of a two-dimensional quantum well (meaning without in-plane electric fields), we now look at the SO effects on the PSB in a quantum dot. From the device geometry, we expect that the interdot tunneling happens along the $x$-direction and thus is introduced only by the $k_x$ operator. We therefore set, apart from $k_y$ and $E_x$, also $k_z$ to zero in Eq. (6) and obtain the SO Hamiltonian for the dot,

$$H_{SO}^{PSB} = k_x\bigl(c_1 E_y \sigma_z + c_3 E_z \sigma_y\bigr). \tag{9}$$

Identifying $\vec{k}_{12}$ with $k_x$ gives Eq. (1) with the explicit form for the spin-orbit field: $\vec{M}_{SO}$ is expected to be in the $y$ - $z$ plane. This is in good agreement with our measurements noting that $(\phi_{SO}, \theta_{SO}) = (0.496\pi, 0.432\pi)$ corresponds to $\vec{M}_{SO} \propto (0.01, 0.98, 0.21)$, almost fully in the $y$-$z$ plane (the deviation is $0.004\pi$ or $0.7°$). In our sample, $\vec{M}_{SO}$ makes an angle of $0.068\pi$ ($12.3°$) with the $y$-axis [Footnote6]. From that result we have a rough estimate on the ratio of the constants $c_1$ and $c_3$, noting that we obtained $\frac{c_1 E_y}{c_3 E_z} \approx 5$. While it is difficult to assign values for the electric fields, based on the gate geometry and applied voltages we speculate that the electric fields along z and y are comparable. It would imply that the constant $c_1$ is bigger than the constant $c_3$, by a factor between 1 and 10.

Before concluding, let us make the following comment. We have assumed that the DQD is described by uniform spin-orbit and g-factor parameters. Since they are probably strongly affected by applied electric fields, it is legitimate to consider that these parameters are actually spatially inhomogeneous: in the next level of model refinement, each dot would be assigned its own spin-orbit field and/or g-factor [68], then perhaps the barrier, too, and so on, until a fully atomistic description [69]. Given the number of uncertainties in our experiment, we feel that the simplest model is adequate as the first approach.

## V.     CONCLUSIONS

We determined the direction of the spin-orbit field $\vec{M}_{SO}$ from the angle dependence of the PSB leakage current in a physically defined Si p-type MOS DQD on silicon-on-insulator. We also identified the related direction for the external magnetic field $\vec{B}_{ext,\|}$ which minimizes the spin-orbit effects on PSB. The measured spin-orbit field is in good accordance with the theory relying only on the crystal structure, the wafer orientation and the device geometry. By orienting the external magnetic field along $\vec{B}_{ext,\|}$ and reducing the background current further by tuning the tunneling rates (inaccessible in this study due to device instability), we will be able to perform PSB spin readout necessary for spin manipulation measurements. Our results may help to devise ways to increase the hole spin coherence and controllability in Si structures. Further studies will include also investigations of the anomalous leakage behavior to learn more about the strength of any residual spin-lifting mechanisms.

## ACKNOWLEDGMENTS

We acknowledge fruitful discussions with Marcel Serina. This work was supported financially by Core Research for Evolutional Science and Technology (CREST), Japan Science and Technology Agency (JST) (JPMJCR15N2, JPMJCR1675), the ImPACT Program of Council for Science, Technology and Innovation (Cabinet Office, Government of Japan) , MEXT Quantum Leap Flagship Program (MEXT Q-LEAP) Grant Number JPMXS0118069228, Japan and Japan Society for the Promotion of Science Grants-in-Aid for Scientific Research (JSPS KAKENHI) grant numbers JP26709023, JP16F16806, 26220710, 18H01819 and 19K14640. J.Y., T.N. and T.O. acknowledge support from RIKEN Incentive Research Projects. T.O. acknowledges Grants-in-Aid for Scientific Research (No. 16H00817, 17H05187), PRESTO (JPMJPR16N3), JST, Futaba Electronics Memorial Foundation Research Grant, Kato Foundation for Promotion of Science Research Grant, Hitachi Global Foundation Kurata Grant, The Okawa Foundation for Information and Telecommunications Research Grant, The Nakajima Foundation Research Grant, Japan Prize Foundation Research Grant, Iketani Science and Technology Foundation Research Grant and Yamaguchi Foundation Research Grant. The Mikiya Science and Technology Foundation Research Grant and Harmonic Ito Foundation Research Grant.

## Appendix A: Symmetry analysis

Here we list the group-theory results stemming from the symmetry analysis of the quantum well and the device. We first analyze the (110)-Si/SiO$_2$ quantum well using a procedure inspired by Ref. [53]. Then we use the group theory to derive the spin-orbit- and magnetic-interaction terms allowed in the valence band Hamiltonian. Finally, we filter the list of the derived terms keeping only the in-plane momenta and the electric fields expected in our sample.

### Symmetry analysis of the Si/SiO$_2$ quantum well

Our quantum well is a slab of silicon with [110] growth direction bordered by a pair of interfaces to oxidized silicon. While the silicon dioxide is amorphous, and, therefore, strictly speaking there is no structural symmetry present, [Footnote7] we reinstate it by assuming that the interface potential is self-averaging on the length scale of the quantum dot, much larger than the interatomic distance. In another words, we assume that (on this scale) the oxide side of the interface is uniform and we can restrict ourselves to considering only the silicon interior between the two [110] interfaces. First to note is that an atomically flat [110] surface of the silicon crystal has symmetry $C_{1h}$. "Averaging" this surface over monoatomic steps [67,70,71] along [110] increases the symmetry to $C_{2v}$ [67]. Taking into account the O$_h$ point group of the bulk silicon, we found five symmetry groups [Footnote8] possibly describing our quantum well [Footnote9]:

1. $^xC_{1h}$ if the two oxide sides are non-equivalent and the two interfaces are atomically flat,
2. $^zC_{2v}$ if the two oxide sides are non-equivalent and each of the two interfaces is "averaged",
3. $^xC_{2h}$ if the two oxide sides are equivalent and there are even number of monoatomic planes in the well,
4. $^yC_{2v}$ if the two oxide sides are equivalent and there are odd number of monoatomic planes in the well,
5. $^yD_{2h}$ if the two oxide sides are equivalent and each of the two interfaces is "averaged".

Here the upper index preceding the group denomination denotes the main axis, using $x = [1\bar{1}0]$, $y = [110]$, $z = [001]$. Illustrative atomic arrangements representing the five cases are given in Fig. A1. We use the symmetry group notation of Ref. [72], see App. A therein. We have illustrated these symmetry groups in Fig.

A1.

## Terms allowed in the Hamiltonian

With the symmetry group specified, we derive the terms allowed in the Hamiltonian using the standard representation theory [72]. In this derivation, we assume that:

  1) The quantum-well confinement lifts the heavy hole-light hole degeneracy of the bulk valence band, so that the ground state manifold is, at zero magnetic field, a Kramers degenerate doublet. Being *effectively* a spin ½ system, we can assign to this subspace a vector of spin ½ operators, the Pauli matrices, which we denote as $\{\sigma_x, \sigma_y, \sigma_z\}$. Importantly, this generic assignment covers any degree of light hole-heavy hole mixing, which is supposed to be generally strong in silicon valence band where the heavy hole-light hole offset is rather low (for unstrained samples) [Footnote10].

  2) There are appreciable electric fields since large voltages are applied on nearby gates. While those electric fields break the symmetry of the quantum well, we include them explicitly (rather than implicitly, by lowering the symmetry) as a vector $\{E_x, E_y, E_z\}$.

  3) We are interested in the low-energy Hamiltonian; thus, we perform an expansion in powers of the hole momenta $\{k_x, k_y, k_z\}$.

  4) Possibly, an external magnetic field is applied, a vector with components $\{B_x, B_y, B_z\}$.

We now derive combinations of these four objects, the two vectors $\vec{E}$ and $\vec{k}$ and the two pseudo-vectors $\vec{B}$ and $\vec{\sigma}$ allowed under the five groups above. We obtain:

Spin-orbit terms generated by the interfaces (that is, linear in the momentum and spin) [Footnote11]:

$$H^1_{\text{SO-IIA}} = \vec{k} \cdot \begin{pmatrix} \{\} & \{1\} & \{1\} \\ \{1\} & \{\} & \{\} \\ \{1\} & \{\} & \{\} \end{pmatrix} \cdot \vec{\sigma}^T, \qquad H^2_{\text{SO-IIA}} = \vec{k} \cdot \begin{pmatrix} \{\} & \{\} & \{1\} \\ \{\} & \{\} & \{\} \\ \{1\} & \{\} & \{\} \end{pmatrix} \cdot \vec{\sigma}^T,$$

$$H^{3,5}_{\text{SO-IIA}} = 0, \qquad H^4_{\text{SO-IIA}} = \vec{k} \cdot \begin{pmatrix} \{\} & \{1\} & \{\} \\ \{1\} & \{\} & \{\} \\ \{\} & \{\} & \{\} \end{pmatrix} \cdot \vec{\sigma}^T.$$

(A1)

Spin-orbit terms generated by an electric field (linear in the momentum, spin, and electric field) [Footnote12]:

$$H^{1,3}_{\text{SO-SIA}} = \vec{k} \cdot \begin{pmatrix} \{E_x\} & \{E_y, E_z\} & \{E_y, E_z\} \\ \{E_y, E_z\} & \{E_x\} & \{E_x\} \\ \{E_y, E_z\} & \{E_x\} & \{E_x\} \end{pmatrix} \cdot \vec{\sigma}^T, \qquad H^{2,4,5}_{\text{SO-SIA}} = \vec{k} \cdot \begin{pmatrix} \{\} & \{E_z\} & \{E_y\} \\ \{E_z\} & \{\} & \{E_x\} \\ \{E_y\} & \{E_x\} & \{\} \end{pmatrix} \cdot \vec{\sigma}^T. \quad (A2)$$

Zeeman (or g-tensor) terms (linear in the spin and magnetic field):

$$H_Z^{1,3} = \vec{B} \cdot \begin{pmatrix} \{1\} & \{\} & \{\} \\ \{\} & \{1\} & \{1\} \\ \{\} & \{1\} & \{1\} \end{pmatrix} \cdot \vec{\sigma}^T, \qquad H_Z^{2,4,5} = \vec{B} \cdot \begin{pmatrix} \{1\} & \{\} & \{\} \\ \{\} & \{1\} & \{\} \\ \{\} & \{\} & \{1\} \end{pmatrix} \cdot \vec{\sigma}^T. \qquad (A3)$$

First order g-tensor corrections (linear in the spin, magnetic field and linear in either the momentum or the electric field)

$$H_{Z-I}^1 = \vec{B} \cdot \begin{pmatrix} \{Y,Z\} & \{X\} & \{X\} \\ \{X\} & \{Y,Z\} & \{Y,Z\} \\ \{X\} & \{Y,Z\} & \{Y,Z\} \end{pmatrix} \cdot \vec{\sigma}^T, \qquad H_{Z-I}^2 = \vec{B} \cdot \begin{pmatrix} \{Y\} & \{X\} & \{\} \\ \{X\} & \{Y\} & \{Z\} \\ \{\} & \{Z\} & \{Y\} \end{pmatrix} \cdot \vec{\sigma}^T,$$

$$H_{Z-I}^{3,5} = 0, \qquad H_{Z-I}^4 = \vec{B} \cdot \begin{pmatrix} \{Z\} & \{\} & \{X\} \\ \{\} & \{Z\} & \{Y\} \\ \{X\} & \{Y\} & \{Z\} \end{pmatrix} \cdot \vec{\sigma}^T. \qquad (A4)$$

Second order g-tensor corrections (linear in spin, magnetic field and quadratic in, counted together, the momentum and the electric field)

$$H_{Z-II}^{1,3} = \vec{k} \cdot \begin{pmatrix} \{X^2,Y^2,Z^2,YZ\} & \{XY,XZ\} & \{XY,XZ\} \\ \{XY,XZ\} & \{X^2,Y^2,Z^2,YZ\} & \{X^2,Y^2,Z^2,YZ\} \\ \{XY,XZ\} & \{X^2,Y^2,Z^2,YZ\} & \{X^2,Y^2,Z^2,YZ\} \end{pmatrix} \cdot \vec{\sigma}^T,$$

$$H_{Z-II}^{2,4,5} = \vec{k} \cdot \begin{pmatrix} \{X^2,Y^2,Z^2\} & \{XY\} & \{XZ\} \\ \{XY\} & \{X^2,Y^2,Z^2\} & \{YZ\} \\ \{Z\} & \{YZ\} & \{X^2,Y^2,Z^2\} \end{pmatrix} \cdot \vec{\sigma}^T.$$

(A5)

These expressions use the following notation: In the Hamiltonian denomination $H_A^B$ the subscript A is a shorthand description for the interaction type as described in the sentence preceding the equation. The superscript B gives indexes of the groups for which the right hand applies. Each set of terms for a given group and an interaction type is represented as a matrix multiplied (the standard vector and matrix multiplications) by a row vector from the left (either $\vec{B}$ or $\vec{k}$) and the column vector of Pauli matrices from the right. The matrix should be interpreted as follows: each element of a list in each (row and column-specified) matrix element stands for a term which enters the Hamiltonian with its own prefactor (the symmetry analysis cannot reveal its value). For example, $H_{SO-SIA}^{1,3}$ in Eq. (A2) means that for the group number 1 or number 3, there are 13 spin-orbit terms, each with its own prefactor. The first three of these, corresponding to the first two columns in the first row of the matrix are $c_1 k_x E_x \sigma_x + c_2 k_x E_y \sigma_y + c_3 k_x E_z \sigma_y$, and so on. If the list is empty, it means no term is associated, while $\{1\}$ corresponds to a constant (that is, a prefactor only): for example, the Zeeman interaction for group 2 is $H_Z^2 = c_1 B_x \sigma_x + c_2 B_y \sigma_y + c_3 B_z \sigma_y$. Finally, for the g-tensor corrections, we use $X$ for either $k_x$ or $E_x$ (thus $X = \{k_x, E_x\}$ in the notation of Eq. A2) and analogously for other Cartesian coordinates [Footnote13]. A matrix entry such as $XY$ then represents 4 possible terms (again, each with its own prefactor): $k_x k_y, k_x E_y, E_x k_y$, and $E_x E_y$.

**Terms relevant for the PSB experiment**

To apply the above results for the specifics of our experiment, we start with the largest symmetry group $D_{2h}$ (number 5 in the list). Namely, as all remaining groups are its subgroups, one might expect that there is a hierarchy of terms: each time the symmetry is lowered, new terms are generated in the Hamiltonian with prefactors somewhat smaller than those corresponding to the preceding higher symmetry [73]. With this, we get the spin-orbit terms from Eq. (A1) – (A2) as

$$H_{\text{SO}} = c_1 E_y k_x \sigma_z + c_2 E_y k_z \sigma_x + c_3 E_z k_x \sigma_y + c_4 E_x k_z \sigma_y + c_5 E_z k_y \sigma_x + c_6 E_x k_y \sigma_z. \qquad (A6)$$

Restricting the momentum to in-plane, appropriate for quasi-two-dimensional holes, that is, setting $k_y = 0$, Eq. (A6) gives Eq. (6) of the main text. Putting further the electric field along the dot main axis to zero, $E_x = 0$, we obtain Eq. (9) of the main text. We also find that lower symmetry of the interfaces (we take C1 for illustration) would induce additional terms (here we have already put $E_x$ and $k_y$ to zero)

$$H_{\text{SO}} = (c_7 E_y + c_8) k_x \sigma_y + (c_9 E_z + c_{10}) k_x \sigma_z + (c_{11} E_z + c_{12}) k_z \sigma_x. \qquad (A7)$$

We analyze the magnetic field interaction analogously. Equation (A3) gives the following leading terms

$$H_Z = g_1 B_x \sigma_x + g_2 B_y \sigma_y + g_3 B_z \sigma_z, \qquad (A8)$$

being Eq. (5) of the main text upon setting $g_1/g_3 = 2$. A lower symmetry [see Eq. (A3)] or higher order contributions [see Eq. (A4) – (A7)] allow for a non-zero $g_{yz}$ term in the g-tensor, reproducing Eq. (4) of the main text. It is interesting to note that (while keeping $E_x = 0$ and $k_y = 0$) none of the terms in Eq. (A4)-(A7) contains a nonzero contribution for the elements of the g-tensor which are zero in Eq. (4) of the main text. We therefore expect the form of the g-tensor in Eq. (4) to be robust.

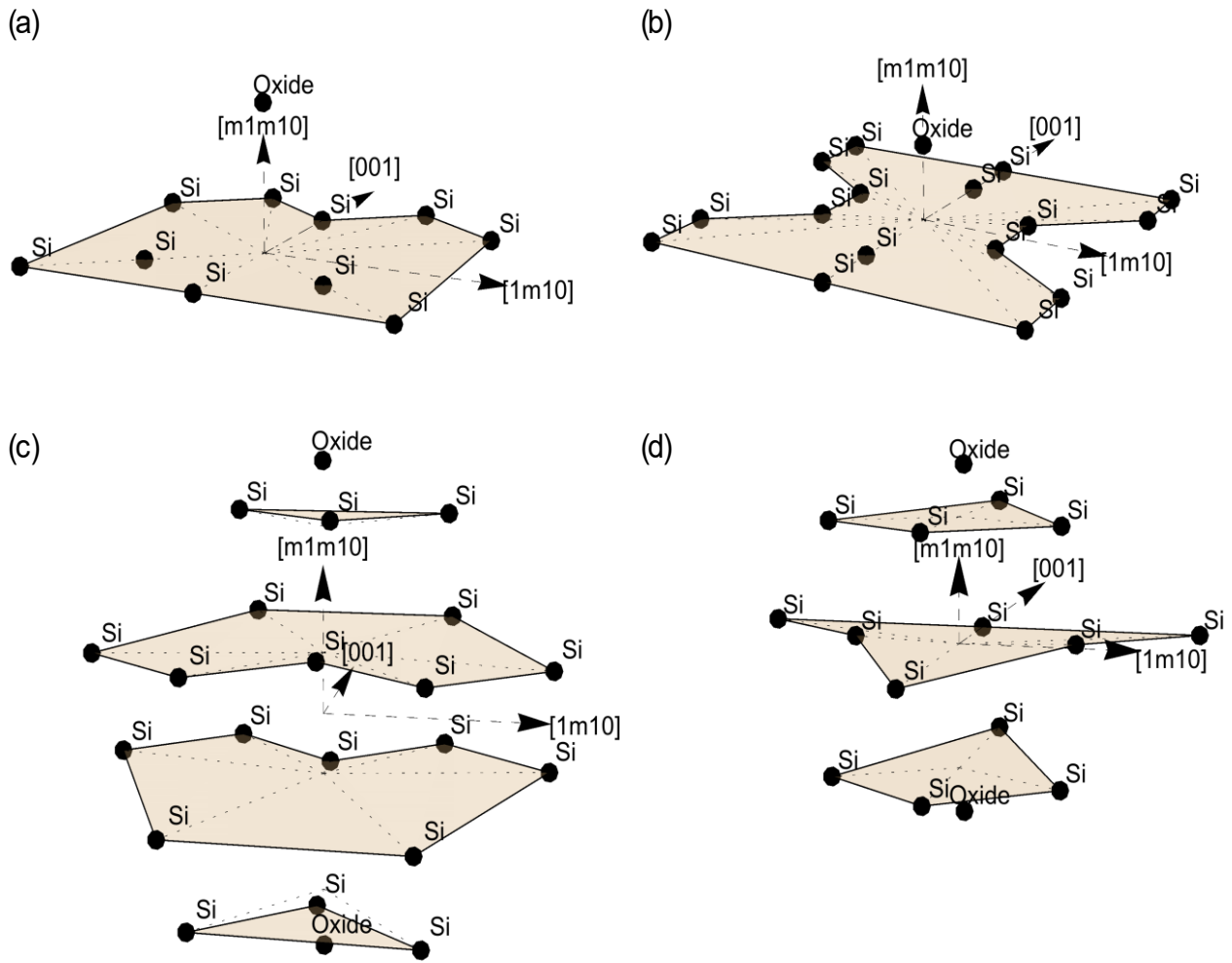

**Figure A1. Illustrative atomic arrangements of the quantum well or its interface plane. (a) Interface plane with an oxide on one side. The group symmetry is $C_{1h}$, with one reflection plane. (b) An "averaged" interface plane (obtained by merging two monoatomic planes) with an oxide on one side. There are two reflection planes and a two-fold rotation axis, resulting in group $C_{2v}$. (c) Four (representing an even number of) monoatomic planes with oxide on both sides. With the center of coordinate system between the planes, there is a two-fold rotation axis, reflection plane and an inversion, group $C_{2h}$. (d) Three (representing an odd number of) monoatomic planes with oxide on both sides. With the center of coordinate system inside the middle plane, there is a two-fold rotation axis and two reflection planes, group $C_{2v}$. The fifth group would correspond to having the oxide also on the bottom side of the structure in panel (b), what adds the inversion symmetry and changes the group to $D_{2d}$.**

## Appendix B: Conversion of $\chi^2$ to confidence intervals

We remind the reader that the minimization of $\chi^2$ is a standard method to find the best fit for a given data set as described by Eq. (B1) [74]:

$$\chi^2 = \sum_{i=0}^{N-1} \left(\frac{y_i - f(x_i)}{\sigma_i}\right)^2. \tag{B1}$$

Here, $(x_i, y_i)$ denote the *ith* data point, with $\sigma_i$ the corresponding standard deviation and $f$ being the tested fitting function.

It is well known [74] that obtaining confidence intervals from the least square fit statistics ($\chi^2$), rests on rather stringent conditions (of errors having normal distributions) which are seldom fulfilled in practice. We, therefore, rely chiefly on $\chi^2$ in the main text. Nevertheless, *for illustration*, we have converted the plotted $\chi^2$ contours also to confidence intervals *assuming* that the errors are normal. Here, we give two formulas useful for the conversion, both taken from Ref. [74]. The first one gives the probability to get $\chi^2_{\min}$ to be $x$ or larger,

$$p(\chi^2_{\min} \geq x \mid N, M) = Q\left(\frac{N-M}{2}, \frac{x}{2}\right). \tag{B2}$$

The second formula gives the probability of the true parameters expressed through the deviation of their $\chi^2$ value from the minimum, $\chi^2_{\min}$, being

$$p(\chi^2 - \chi^2_{\min} \geq x \mid N, M) = Q\left(\frac{M}{2}, \frac{x}{2}\right). \tag{B3}$$

In these formulas, $N$ is the number of data points, $M$ is the number of fitting parameters, and $Q$ is the incomplete Gamma function,

$$Q\left(z, \frac{x}{2}\right) = \frac{1}{\int_0^\infty t^{z-1} e^{-t} dt} \int_{\frac{x}{2}}^\infty t^{z-1} e^{-t} dt \tag{B4}$$

For the in-plane data, we have $N = 32$ data points, $M = 3$ fitting parameters, and the best fit with $\chi^2_{\min} = 58.5$. Assuming normal errors, such a value would be obtained with probability $p(\chi^2_{\min} \geq 58.5) \approx 10^{-3}$. This is a tolerable fit goodness and the conversion of $\chi^2$ to the confidence region in Fig.2(c) is reasonably faithful. The conversion of $p \in \{68\%, 95\%, 99.7\%\}$ using Eq. (B3) gives $\chi^2 - \chi^2_{\min} \in \{3.5, 8, 14\}$, respectively.

For the in-plane and out-of-plane data combined, we have $N = 51$, $M = 4$, and $\chi^2_{\min} = 164.4$. The associated probability is $p(\chi^2_{\min} \geq 164) \approx 10^{-14}$. Most probably, this low fit goodness is due to underestimating the errors on the data plotted in Fig. 2(f). It means that the assignment, done using Eq. (B3) converting $p = 99.7\%$ to $\chi^2 - \chi^2_{\min} \approx 16$, of the "$3\sigma$" tag for contour plotted in Fig. 2(g) is not very reliable.

[Footnote1] The dependence of the leakage current on the detuning $\varepsilon$ was reported in Ref. [46].

[Footnote2] The same idea has been pursued in III-V element nanowire dots in Ref. [41] for holes and Ref. [75] for electrons.

[Footnote3] Compared to electrons, the PSB lifting of holes is on one hand simpler due to the absence of nuclear spin effects [76], on the other hand more complicated due to the g-factor anisotropy [77].

[Footnote4] We still need to fix the relative signs of the elements. Based on k.p estimates, see, for example, Ref. [78], we choose all elements in Eq. (5) as positive.

[Footnote5] The names are motivated by the dependence of the effective magnetic field vector direction on the in-plane momentum direction. In this respect, the two bottom panels in Fig. 3 correspond to the well-known Rashba and Dresselhaus terms of the GaAs conduction band. We emphasize that here both terms are induced by external electric fields, so from this point of view one could also call both of them Rashba-like. The names we use are in line with Ref. [53].

[Footnote6]: Given a sizable uncertainty in the extracted SO field, a deflection out of the $y$-$z$ plane cannot be excluded. The most straightforward way to induce such a component is to consider that the double dot axis is not completely aligned with the $x$-axis, so that $\vec{k}_{12}$ would contain a finite $k_z$ component. Equation (6) then gives in a term proportional to $\sigma_x$.

[Footnote7] The structure and build-up of the Si-SiO$_2$ interface was investigated extensively, see Ref. [79], and the references therein, especially Refs. [80–82].

[Footnote8] We do not include explicitly the possibility of no spatial symmetry which allows for any term in the Hamiltonian.

[Footnote9] These five cases, in the order given, correspond to the five cases for a Si/SiGe [001] quantum well investigated in Ref. [53] in the following order: $C_{2v}, C_{4v}, D_{2h}, D_{2d}, D_{4h}$.

[Footnote10] In unstrained Si, the ground state is supposed to be a heavy hole in two-dimensional geometries [83], and a light hole in one-dimensional ones (nanowires) [84]. From among the most common growth directions of [001], [110], and [111], the heavy-hole-light-hole splitting is largest for [110] [83]. For example, Ref. [85] measured the splitting of about 40 meV. However, strain has strong effects [86–88] and can change the splitting appreciably even inverting the states' ordering.

[Footnote11] In the nomenclature of Ref. [19], these terms would be called "interface inversion asymmetry (IIA)" origin. Refs. [54,89] are two examples which derive such terms from a microscopic model of the interface.

[Footnote12] These terms would be called of "structure inversion asymmetry (SIA)" origin.

[Footnote13] The equivalence between momentum and electric field follows since both are vectors which transform identically under the five groups and are therefore indistinguishable from the symmetry point of view.